\documentclass[prl,superscriptaddress,amsmath,amssymb,floatfix,showpacs,twocolumn]{revtex4-1}

\usepackage{epsfig}
\usepackage{times}
\usepackage{graphicx}
\usepackage{bm} % bold maths, for all characters, including greek
\usepackage{color}
\usepackage{dcolumn} % align decimal points
\usepackage{array}
\usepackage{rotating} % for sideways tables, figures, etc.
\usepackage{longtable} % for long tables
\newcommand{\ave}[1]{\langle #1 \rangle}

\newcommand{\s}{\sigma}

\newcommand{\eg}{e_{g}}
\newcommand{\ttg}{t_{2g}}
\newcolumntype{d}{D{.}{.}{1.2}}

\begin{document}

\title{Towards a first-principles determination of effective
Coulomb interactions in correlated electron materials:
Role of intershell interactions}

\author{Priyanka Seth}
\affiliation{Centre de Physique Th\'eorique, Ecole Polytechnique,
CNRS UMR 7644, 91128 Palaiseau, France}

\author{Philipp Hansmann}
\affiliation{Centre de Physique Th\'eorique, Ecole Polytechnique,
CNRS UMR 7644, 91128 Palaiseau, France}
\affiliation {Max-Planck-Institut f\"ur Festk\"orperforschung,
Heisenbergstrasse 1, 70569 Stuttgart, Germany}

\author{Ambroise van Roekeghem}
\affiliation{Centre de Physique Th\'eorique, Ecole Polytechnique,
CNRS UMR 7644, 91128 Palaiseau, France}
\affiliation{Beijing National Laboratory for Condensed Matter Physics,
and Institute of Physics, Chinese Academy of Sciences, Beijing 100190, China}

\author{Loig Vaugier}
\affiliation{Centre de Physique Th\'eorique, Ecole Polytechnique,
CNRS UMR 7644, 91128 Palaiseau, France}

\author{Silke Biermann}
\affiliation{Centre de Physique Th\'eorique, Ecole Polytechnique,
CNRS UMR 7644, 91128 Palaiseau, France}

\date{\today}

%%%%%%%%%%%%%%%%%%%%%%%%%%%%%%%%%%%%%%%%%%%%%%%%%%%%%%%%%%%%%%%%%%%%%%%%%%%%%%%%
\begin{abstract}
The determination of the effective Coulomb interactions to be used in
low-energy Hamiltonians for materials with strong electronic correlations
remains one of the bottlenecks for parameter-free electronic structure
calculations.  We propose and benchmark a scheme for determining the effective
local Coulomb interactions for charge-transfer oxides and related compounds.
Intershell interactions between electrons in the correlated shell and ligand
orbitals are taken into account in an effective manner, leading to a reduction
of the effective local interactions on the correlated shell.  Our scheme
resolves inconsistencies in the determination of effective interactions as
obtained by standard methods for a wide range of materials, and allows for a
conceptual understanding of the relation of cluster model and dynamical mean
field-based electronic structure calculations.
\end{abstract}

\pacs{71.15-m,71.27.+a,71.10.Fd}

\maketitle

\newpage

%%%%%%%%%%%%%%%%%%%%%%%%%%%%%%%%%%%%%%%%%%%%%%%%%%%%%%%%%%%%%%%%%%%%%%%%%%%%%%%%
%%%%%%%%%%%%%%%%%%%%%%%%%%%%%%%%%%%%%%%%%%%%%%%%%%%%%%%%%%%%%%%%%%%%%%%%%%%%%%%%
%%%%%%%%%%%%%%%%%%%%%%%%%%%%%%%%%%%%%%%%%%%%%%%%%%%%%%%%%%%%%%%%%%%%%%%%%%%%%%%%

The behavior of electrons in the immediate proximity of the Fermi level
determines most of the interesting response properties of materials with strong
electronic Coulomb correlations. The {\it ab initio} derivation of effective
Hamiltonians that capture this low-energy physics has therefore become a
crucial milestone of modern condensed matter theory, and a prerequisite for
materials-specific many-body calculations.  While density functional techniques
(DFT) \cite{kohn_nobel_1999} provide an established tool to construct the
one-body part of the low-energy Hamiltonian, ``downfolding'' the interaction
part is more subtle.  The recent development of constrained screening
approaches, such as the constrained random phase approximation (cRPA), that
identify the interaction parameters to be used in subsequent many-body
calculations in a low-energy subspace as matrix elements of partially screened
Coulomb interactions
\cite{aryasetiawan_frequency-dependent_2004,vaugier_hubbard_2012}, have led to
tremendous progress, but have also raised new questions.  In several systems of
recent interest such as atoms adsorbed on surfaces
\cite{hansmann_long-range_2013, hansmann_what_2013} or graphene
\cite{schuler_optimal_2013}, intersite Coulomb interactions, neglected in the
Hubbard model, were found to reach values of up to 50$\%$ of the local
interactions, triggering work on the inclusion of long-range effects
\cite{ayral_spectral_2012,ayral_screening_2013,huang_extended_2014,
mahmoudian_glassy_2015}.

In this Letter, we demonstrate that in prototypical correlated transition metal
(TM) oxides and $f$-electron systems, standard calculations omit an even more
important additional term, namely intershell interactions that couple the
``correlated'' shell to itinerant states. Examples are the $d$-$f$ interaction
in $f$-electron materials or $d$-ligand interactions in TM oxides.  We evaluate
these interactions from first principles and derive a scheme to determine {\it
effective intrashell interactions}, which are renormalized by intershell
interactions.  Our calculations for various materials including Ce, NiO and
$f$-electron insulators demonstrate that this ``shell-folding scheme'' not only
fixes present problems in determining effective interactions from first
principles, but also gives new insight in how to treat late TM oxides within
dynamical mean-field-based (DMFT) techniques
\cite{wang_covalency_2012_han_dynamical_2011,haule_covalency_2014}.

We begin by calculating, using the cRPA implementation of
\cite{vaugier_hubbard_2012}, the interactions between TM $3d$ (or actinide
$4f$) orbitals and O $2p$ ligands to be used in a low-energy Hamiltonian
treating both orbital species as correlated.  For the actinide oxide UO$_2$ we
find $U^{fp}=1.9$ eV, as compared to an onsite $f$-$f$ interaction of 6.5 eV
and an intersite $V^{ff}$ of 1 eV.  For NiO, $U^{dp} = 2.2$ eV still exceeds
25$\%$ of the onsite $U^{dd} = 8.6$ eV, and is larger than $V^{dd} = 1.5$ eV.
These findings might explain the difficulties that DMFT-based electronic
structure calculations face when dealing with late TM oxides
\cite{karolak_double_2010,wang_covalency_2012_han_dynamical_2011,haule_covalency_2014}
using a Hamiltonian in which intershell interactions are neglected.
Specifically, for NiO, it has been demonstrated that none of the standard
double counting correction forms provide spectral properties in agreement with
experiment \cite{karolak_double_2010}.  The situation is similar for rare-earth
nickelates and high-T$_c$ cuprates for which intershell $d$-$p$ interaction
effects have either been mimicked by adjustable shifts
\cite{kent_combined_2008,weber_optical_2008} or been taken on a static
mean-field level \cite{hansmann_importance_2014}.

Electronic structure calculations for elemental cerium usually assume a Hubbard
$U$ of the order of 5-6 eV to be applied to the $f$-states. For example, DMFT
calculations performed for a low-energy ``$f$-$spdf$'' Hamiltonian comprising
$s$-, $p$-, $d$- and $f$-states, and a Hubbard correction for the $f$-states of
this order of magnitude yield good agreement between theory and a vast body of
experimental probes
\cite{held_cerium_2001,zolfl_spectral_2001,haule_cerium_2005,amadon_cerium_2006,
streltsov_magnetic_2012,bieder_thermodynamics_2014,lanata_interplay_2014,
chakrabarti_alpha-gamma_2014}.
Quite embarrassingly, however, cRPA calculations for an $f$-$spdf$ Hamiltonian,
where localized states are constructed for an energy window containing $s$-,
$p$-, $d$- and $f$-states but the polarization is constrained to only exclude
screening among the $f$-states, give a ridiculously small value ($<$ 1 eV)
\cite{pavarini_lda+dmft_2011,amadon_screened_2014} \footnote{ Interestingly,
even when additional screening processes are cut out, corresponding e.g., to
transitions involving the Ce-t$_{2g}$ states \cite{amadon_screened_2014}, or
when the hybridization is effectively reduced by using a disentangled band
structure \cite{nilsson_ab_2013}, obtained $U$ values remain on the small
side.}. 

\begin{figure}[htpb]
  \centering
    \includegraphics[width=1.00\columnwidth]{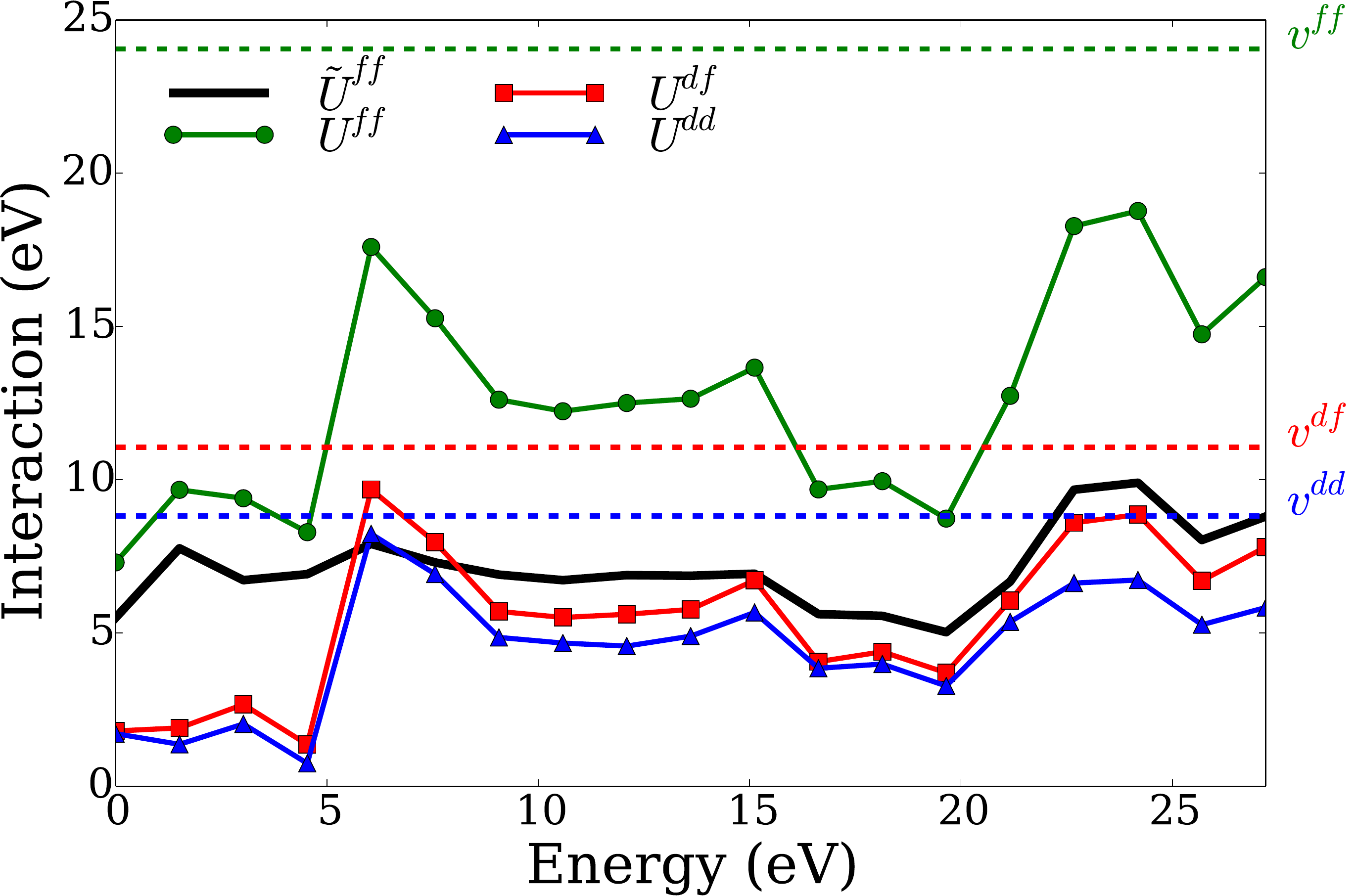}
    \caption{(Color online) ${\cal U}(\omega)$ for $df$-$df$ and effective (SF) Hamiltonians for cerium.
             \label{fig:ce}}
\end{figure}

We argue here that both considering a second shell as correlated and treating
the intershell interaction at least in an effective manner are crucial
ingredients to remedy this problem.  For $\alpha$-Ce, assuming $f$- and
$d$-states to be correlated within cRPA, we find an intershell interaction
$U^{fd}=1.8$ eV, as compared to $U^{ff}=6.5$ eV and a negligible intersite
$V^{ff}=0.01$ eV.  The large difference in the onsite $U^{ff}$ depending on
whether a second shell is treated as correlated (and thus excluded as a
screening channel in the cRPA) can be traced to strong $f$-$spd$ hybridization.
Furthermore, it has been recently argued
\cite{springer_frequency-dependent_1998,aryasetiawan_frequency-dependent_2004,
casula_dynamical_2012,biermann_dynamical_2014} that the effective local Hubbard
interaction is a dynamical quantity. The cRPA allows for a direct assessment of
this frequency-dependence and its consequences have been intensively studied
within DMFT
\cite{casula_dynamical_2012,werner_satellites_2012,huang_dynamical_2012,
casula_low-energy_2012,van_roekeghem_dynamical_2014,krivenko_slave_2015}. We
have calculated ${\cal U}(\omega)$ for for Ce constructing a Hamiltonian that
comprises both $4f$ and $5d$ orbitals as correlated states. The results (see
Fig.~\ref{fig:ce}) display a striking similarity in the frequency dependence of
$U^{ff}$, $U^{fd}$ and $U^{dd}$, and in the magnitude of $U^{fd}$ and $U^{dd}$,
suggesting that screening is dominantly due to the same screening processes
(e.g., transitions from $f$- or $d$-states to $sp$-dominated states via strong
hybridization as well as plasmon-like excitations).  As a consequence, the
differences $\tilde{U}^{ff/dd} = U^{ff/dd} - U^{fd}$ are only weakly
frequency-dependent quantities. Below, we will argue that these differences
acquire a physical meaning as effective interactions when intershell
interactions are taken into account in an effective manner.

Consider a multi-orbital Hubbard Hamiltonian for materials where two different
orbital shells are treated as correlated.  Since, besides Ce, important
examples are provided by the TM $d$- and ligand $p$-states in TM compounds, we
will refer to these two shells as $d$- and $p$-shells, with the respective
``correlated'' and ``ligand'' subspaces denoted by ${\cal C}$ and ${\cal L}$.
Our basic assumption is that DFT, within the local density (LDA) or generalized
gradient approximation, gives a good estimate for the total energy as a
function of the expectation values of the number operators $N_d$ and $N_p$ of
the two shells.  We further assume that we know the intra- and intershell
interactions $U^{dd}$, $U^{pp}$, and $U^{dp}$.  In practice, we obtain those
from the standard cRPA by suppressing {\it all} screening channels within the
full ${\cal C}$+${\cal L}$ space.  Following the spirit of ``LDA++''
\cite{lichtenstein_abinitio_1998,anisimov_first-principles_1997}, we can then
define a {\it bare} one-body Hamiltonian $H_0$ as the LDA estimate from which
the average interactions within the $d$- and $p$-shells have been subtracted.
Our multi-orbital Hubbard Hamiltonian now reads: $H = H_0 + H_{\rm int} - \mu
N_{\rm tot}$, where $H_{\rm int} = \sum_{\bf R} h_{\bf R}$ becomes a sum over
the positions ${\bf R}$ of the atoms carrying the correlated shell, comprising
intra- and intershell Hubbard terms:
\begin{eqnarray}
\label{eq:general-pd-H}
h_{\bf R} &=& 
\frac{1}{2} \sum_{\begin{subarray}{c}
                   (m\s) \ne (m'\s^{\prime})\\  
                   m,m' \in {\cal C}
                  \end{subarray}}
U^{dd}_{m \s m' \s'} n_{m\s} n_{m'\s^{\prime}} 
\\
&+&
\frac{1}{2} \sum_{\begin{subarray}{c}
                   (m\s) \ne (m'\s^{\prime})\\  
                   m,m' \in {\cal L}
                  \end{subarray}}
U^{pp}_{m \s m' \s'} n_{m\s} n_{m'\s^{\prime}}
+ 
\sum_{\s,\s'}
U^{dp} N_{d \s} N_{p \s'}.
\nonumber
\end{eqnarray}
Here, $U^{dd}_{m \s m' \s'}$ and $U^{pp}_{m \s m' \s'}$ are the spin- and
orbital-resolved interaction matrix elements, 
$N_{(d/p) \s} = \sum_{\begin{subarray}{c}
                    m \in {\cal C}/{\cal L}
                   \end{subarray}} n_{m\sigma}$
and 
$n_{m \sigma} = c^{\dagger}_{m \sigma} c_{m \sigma}$.
To simplify the notation, we have omitted the ${\bf R}$ indices on the
operators, and the intershell interactions couple only to the total charge on a
given shell.  A purely algebraic manipulation allows us to rewrite this
Hamiltonian as
\begin{eqnarray}
\label{eq:eff-pd-H}
h_{\bf R} &=& 
\frac{1}{2} \sum_{\begin{subarray}{c}
                   (m\s) \ne (m'\s^{\prime})\\  
                   m,m' \in {\cal C}
                  \end{subarray}}
\tilde{U}_{m \s m' \s'}^{dd} n_{m\s} n_{m'\s^{\prime}}
\\
&+&
\frac{1}{2} \sum_{\begin{subarray}{c}
                   (m\s) \ne (m'\s^{\prime})\\  
                   m,m' \in {\cal L}
                  \end{subarray}}
\tilde{U}_{m \s m' \s'}^{pp} n_{m\s} n_{m'\s^{\prime}}
+
\frac{U^{dp}}{2} N (N - 1),
\nonumber
\end{eqnarray}
with $N = \sum_{\s} (N_{d \s} + N_{p \s})$ and 
$\tilde{U}_{m \s m' \s'} = U_{m \s m' \s'}-U^{dp}$.

The key observation here is that a Hamiltonian with explicit intershell
interactions can be rewritten as a sum of interacting Hamiltonians for the two
shells, coupled only through an interaction acting on the total charge.  The
calculations further simplify if for some reason the interactions on the
$p$-shell can be neglected. The usual argument consists in invoking the
nearly-complete filling of this shell, which impedes charge fluctuations and
allows replacement of operators $n_p$ by their static mean-field values
$\langle n_p \rangle$, but this is difficult to justify in late TM oxides where
$d$-$p$ fluctuations are generally the most efficient screening mechanism,
substantially reducing the $p$-occupations.  However, as we will see
below, $U^{dp}$ can be comparable in magnitude to the intra-ligand-shell
interactions $U^{pp}$, making the effective interaction $\tilde{U}^{pp} =
U^{pp}-U^{dp}$ small.  The Hamiltonian then has a
$d$-$dp$ form, with correlated $d$ and non-correlated $p$ electrons,
and contains the renormalization of the $dd$-interactions by the intershell
interactions: $\tilde{U}^{dd} = U^{dd}-U^{dp}$.  Analogously to ``downfolding''
in energy space, we refer to this scheme as ``shell-folding'' (SF).

While very suggestive in the sense of the above interpretation,
Eq.~(\ref{eq:eff-pd-H}) involves a certain number of subtleties.  First, we
have divided the solid into cells centred on atoms carrying correlated shells,
and constructed ligand Wannier functions {\it centred on these correlated
atoms}. For the prototypical material NiO, this construction is explicitly
discussed in the Supplemental Material \cite{supplementary}.  It is similar to
what is routinely done in cluster model calculations, where hybridizing
cluster ligand orbitals are combinations of neighbouring ligand orbitals that
are able, by symmetry considerations, to couple to the correlated orbitals.
Akin to a Zhang-Rice-type construction \cite{zhang_effective_1988}, a unitary
transformation rotates the original Hamiltonian in ligand-centred $p$-orbitals
basis into a ``cell Hamiltonian'' expressed in TM-centred ligand Wannier
functions. The above $p$ degrees of freedom are to be understood in this sense
\footnote{The effective $p$-$d$ interaction $U^{dp}$ in this basis can be
expressed in terms of $U^{dp}$ in the ligand-centred basis, and it has been
shown \cite{feiner_effective_1996} that the two coincide up to a numerical
prefactor that is close to 1. For the sake of simplicity, we assume the
prefactor to be equal to 1 here.}.  Furthermore, we have implicitly assumed
that the total particle number {\it on a given cell} is a conserved quantity:
the dominant screening channels involve the immediate neighbour ligands, while
long-range processes within the $dp$-subspace are neglected.  This assumption
is made in cluster model calculations too, and the success of such calculations
can be inferred as an {\it a posteriori} justification of its validity. 

The exact rewriting of the Hamiltonian can be understood in a complementary
manner: one can formulate the problem as the quest for an optimal auxiliary
system with decoupled correlated shells reproducing as closely as possible the
physics of the full system. If the free energy is used as a measure of the
quality of the approximation, such an optimal system can be constructed via the
Peierls-Feynman-Bogoliubov variational principle.
Consider the auxiliary Hamiltonian:
\begin{align*} 
 \tilde{h}_{\bf R} = & \frac{\tilde{U}^{dd}}{2} \sum_{\begin{subarray}{c} 
                                                       (m\s) \ne (m'\s^{\prime})\\
                                                       m,m' \in {\cal C}
                                                      \end{subarray}}
                                            \hspace{-.4cm} 
n_{m \s} n_{m' \s'} 
+ \frac{\tilde{U}^{pp}}{2} \sum_{\begin{subarray}{c} 
                                                       (m\s) \ne (m'\s^{\prime})\\
                                                       m,m' \in {\cal L}
                                         \end{subarray}}
                                            \hspace{-.4cm} 
                                                      n_{m \s} n_{m' \s'}.
\end{align*}
The optimal value of the (for simplicity, spherically- and spin- averaged)
interactions $\tilde{U}^{dd}$ and $\tilde{U}^{pp}$ are those that minimize the
auxiliary free energy $\tilde{F} = F[\tilde{h}_{\bf R}] + \ave{h_{\bf R} -
\tilde{h}_{\bf R}}_0$, where $\ave{\ldots}_0$ refers to the average taken with
the auxiliary Hamiltonian, and $h_{\bf R}$ is given by
Eq.~\ref{eq:general-pd-H}.  From the stationarity condition $ \partial
\tilde{F} / \partial \tilde{U}^{dd} = 0 $ we obtain
\begin{align}
\label{eq:U0_shellfolded}
\tilde{U}^{dd} = U^{dd} + 
\frac{ 2 U^{dp} D^{dp}
       + (U^{pp} - \tilde{U}^{pp}) D^{pp}
      }{D^{dd}},
\end{align}
where (for $X,Y=d$ or $p$)
\begin{eqnarray}
D^{XY} = \frac{\partial}{\partial \tilde{U}^{dd}}
         \sum_{(i\s) \neq (j\s')} \ave{ n^{X}_{i \s} n^{Y}_{j \s'}}_0, 
\end{eqnarray}
and an analogous equation for $\tilde{U}^{pp}$. The optimal interactions are
the solutions of this system of equations.  If, in addition, we assume that $N
= N_{d} + N_{p}$ is invariant with respect to changes in $\tilde{U}^{dd/pp}$,
these solutions simplify to $ \tilde{U}^{dd} = U^{dd} - U^{dp}$ and $
\tilde{U}^{pp} = U^{pp} - U^{dp}$, in agreement with Eq.~(\ref{eq:eff-pd-H})
\cite{supplementary}.  This requirement imposes that charge may be displaced
from the correlated site to the ligand orbitals within the cluster as a result
of changes in the interaction, which is equivalent to our assumption that
screening takes dominantly place within the immediate neighbourhood of a given
atom.  Imposing the requirement that increasing the interaction strength only
results in charge flow from a site to its nearest neighbours in the
single-orbital extended Hubbard model (that is, in an effective Hamiltonian
with local and intersite interactions), gives rise to an analogous result as
derived in \cite{schuler_optimal_2013}. In this case the effective interaction
becomes $\tilde{U} = U - V_{nn}$, where $V_{nn}$ is the nearest-neighbour
interaction. Here, a multi-orbital model with two correlated shells and
intershell interactions is mapped onto a model with reduced intrashell
interactions only and a coupling only to the total charge on a cell, or if
$U^{pp} - U^{dp}$ can also be neglected, onto a model with only one correlated
shell.

\begin{table}[htdp]
\caption{Slater integrals
for correlated shells, 
average intraorbital $U^{pp}$ for 
 ligand $p$-shells and intershell $U^{dp/fp}$
for 
 (a) Ce for $f$-$spdf$, $df$-$df$, $spdf$-$spdf$,
 (b) NiO and actinide oxides for $d$-$dp$, $dp$-$dp$, $f$-$fp$, $fp$-$fp$, and
     SF models as appropriate.
 All values in eV.}
\label{tab:uvalues}
\centering
\scalebox{0.8}{
\begin{tabular}{@{\extracolsep{5pt}}cdddddddddddddd@{}}
{\bf Ce}
 & \multicolumn{1}{c}{$f$-$spdf$}
 & \multicolumn{2}{c}{$df$-$df$}
 & \multicolumn{4}{c}{$spdf$-$spdf$} \\
 & & \multicolumn{1}{c}{$f$-$f$} & \multicolumn{1}{c}{$d$-$d$}
   & \multicolumn{1}{c}{$f$-$f$} & \multicolumn{1}{c}{$d$-$d$} & \multicolumn{1}{c}{$p$-$p$} & \multicolumn{1}{c}{$s$-$s$} \\
\cline{3-4} \cline{5-8}
$F^0$    & 0.9 & 6.5 & 1.5 & 7.9 & 1.8 & 1.0 & 0.9 \\ 
$F^2$    & 5.8 & 8.8 & 2.7 & 9.4 & 3.1 & 3.8 &     \\
$F^4$    & 5.6 & 6.4 & 2.5 & 6.5 & 2.6 &     &     \\
$F^6$    & 4.7 & 5.1 &     & 5.1 &     &     &     \\
$U^{fd}$
 & \multicolumn{1}{c}{}
 & \multicolumn{2}{c}{1.8}
 & \multicolumn{2}{c}{2.4} 
\end{tabular}
}
\scalebox{0.8}{
\begin{tabular}{@{\extracolsep{5pt}}cddddddddd@{}}
& \multicolumn{3}{c}{\bf NiO}
& \multicolumn{4}{c}{\bf UO$_2$} & \multicolumn{1}{c}{\bf PuO$_2$} 
& \multicolumn{1}{c}{\bf CmO$_2$} \\
& \multicolumn{1}{c}{$d$-$dp$} 
& \multicolumn{1}{c}{$dp$-$dp$} 
& \multicolumn{1}{c}{SF}
& \multicolumn{1}{c}{$f$-$f$} & \multicolumn{1}{c}{$f$-$fp$} 
& \multicolumn{1}{c}{$fp$-$fp$} & \multicolumn{1}{c}{SF}
& \multicolumn{1}{c}{SF} & \multicolumn{1}{c}{SF} \\
\cline{2-4} \cline{5-8} \cline{9-9} \cline{10-10}
$F^0$       &  6.1  &  8.6  &  6.4  &  3.5  &  3.9  &  6.5  &  4.6  &  5.3  &  5.9  \\
$F^2$       &  9.7  & 10.1  & 10.1  &  5.3  &  5.8  &  6.2  &  6.2  &  7.3  &  8.1  \\
$F^4$       &  6.6  &  6.8  &  6.8  &  4.3  &  4.9  &  5.0  &  5.0  &  5.6  &  6.1  \\
$F^6$       &       &       &       &  3.9  &  4.1  &  4.2  &  4.2  &  4.4  &  4.8  \\
$U^{pp}$    &       &  6.8  &  4.6  &       &       &  6.0  &  4.1  &  4.3  &  4.6  \\
$U^{dp/fp}$ &       &  2.2  &       &       &       &  1.9  &       & (2.0) & (2.1) \\
\end{tabular}
}
\end{table}

We now discuss how the above shell-folding scheme resolves contradictions
arising in modeling specific materials.  For $\alpha$-Ce, our SF scheme finds
$F^0=4.7$ eV, giving $U^{ff}=5.4$ eV.  This corresponds to the value usually
used for cerium.  More interestingly, the intrashell interactions $U^{dd}=1.8$
eV for the $d$-orbitals are by the same mechanism renormalized to
$\tilde{U}^{dd}=0$, since the intershell interaction equals the initial $d$-$d$
interaction within numerical accuracy, see Fig.~\ref{fig:ce} at $\omega=0$.
This justifies the construction of the Hamiltonian for DMFT calculations for
cerium, and in particular to the fact that only $f$-states are treated as
correlated.  Furthermore, they reconcile the cRPA values with $U \sim 5-6$ eV
needed in practice.

For paramagnetic NiO, the $F^0$ for the correlated $d$-subspace in the SF
Hamiltonian (see Table \ref{tab:uvalues}) is found to be larger than that
calculated within the $d$-$dp$ setup. Extensive cluster model calculations
exist for this compound, where interaction strengths have been adjusted so as
to reproduce experimental spectra.  These optimal values coincide with our
calculations within numerical accuracy: Ref.~\onlinecite{haupricht_local_2012}
uses $F^0=6.5$ eV and finds excellent agreement with experiment (see Fig.~3 in
Ref.~\onlinecite{haupricht_local_2012}).

\begin{figure}[htpb]
  \centering
    \includegraphics[width=1.00\columnwidth]{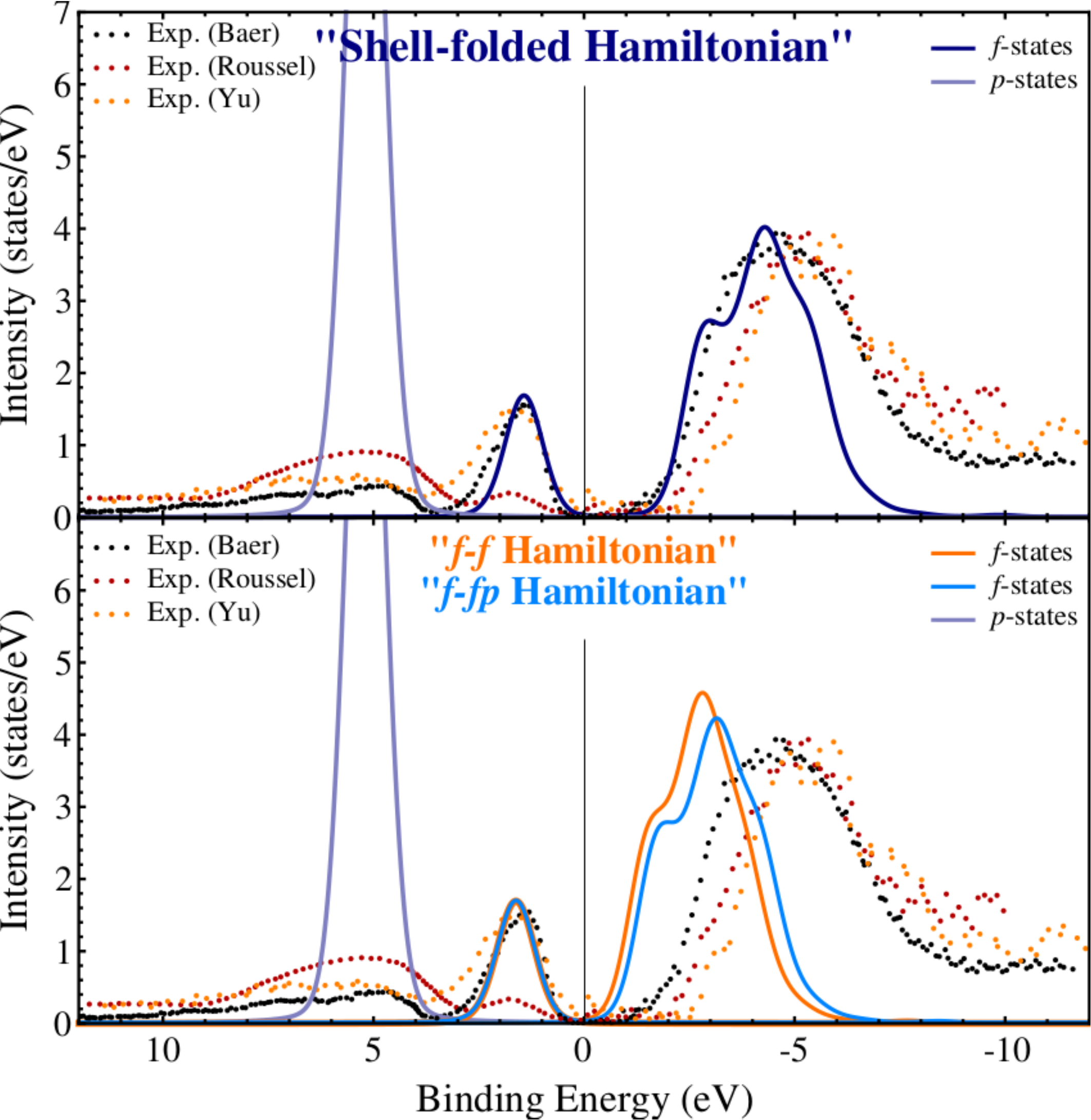}
    \caption{(Color online) PES and IPES spectra for UO$_2$ using interaction 
             parameters obtained in the $f$-$f$, $f$-$fp$ and SF
             models compared to experimental spectra
             \cite{baer_electronic_1980,roussel_inverse_2009,yu_f-f_2011}.
             \label{fig:uo2_spectra}}
\end{figure}

The fluorite structure actinide oxides UO$_2$, PuO$_2$, and CmO$_2$ are
insulating $f$-electron compounds that we consider in their non- or
paramagnetic phases \cite{supplementary}. In particular, UO$_2$ has received
much interest given its role as a nuclear fuel, and its electronic structure
has been studied extensively
\cite{dorado_dft+u_2009,yin_origin_2008,yin_electronic_2011,wen_density_2013}.
We have calculated the interactions corresponding to $f$-$f$, $f$-$fp$ and
$fp$-$fp$ Hamiltonians for UO$_2$.  The Slater integrals $F^k$ needed to
parametrize these interactions are given in Table \ref{tab:uvalues}.  The
increased localization of the $f$-orbitals in the $f$-$fp$ model relative to
the $f$-$f$ model is reflected in the larger values of $F^k$.  The magnitude of
$f$-$f$ interactions in UO$_2$ is significantly larger in the $fp$-$fp$ model,
signaling the importance of $p$-to-$f$ screening transitions.  While
integration of the intershell interactions reduces $F^0$ to the SF value of 4.6
eV, this value remains larger than that obtained in the $f$-only models.  As we
proceed along the actinide series, the value of the $f$-$f$ interaction $F^0$
in the $f$-$fp$ model decreases from 3.9 (UO$_2$) to 3.5 (PuO$_2$) to 3.4 eV
(CmO$_2$). The increased screening efficiency of the O 2$p$-states as the
charge-transfer gap shrinks outweighs the decreased $f$-orbital extension as
the nuclear charge increases.  In contrast, the SF $F^0$ increases from 4.6
(UO$_2$) to 5.3 (PuO$_2$) to 5.9 eV (CmO$_2$).  In this case, transitions
between $p$- and $f$-states do not contribute to the polarization, and the
increased localization of $5f$ states determines the overall trend.  For
UO$_2$, the $p$-$f$ charge-transfer energy calculated from the local part of
the DFT Hamiltonian, $\Delta = 5.2$ eV, is larger than the $\tilde{U}^{ff}=4.6$
eV indicating the Mott-Hubbard character of the gap in UO$_2$ (opening between
$f$-dominated states).  On the other hand, PuO$_2$ and CmO$_2$, which
respectively have $\Delta = 4.2$ and $3.4$ eV, are found to be charge-transfer
insulators, with a gap between $f$- and $p$-states, in agreement with
Ref.~\onlinecite{wen_density_2013}.  With our cRPA results we have performed
simulations of (inverse-) photoemission experiments using a
configuration-interaction cluster model including the atomic-multiplet theory
\cite{tanaka_resonant_1994,de_groot_x-ray_1994,theo_thole_1997} for UO$_2$
\cite{supplementary}.  The spectra obtained using parameters from the $f$-$f$,
$f$-$fp$ and shell-folded $fp$-$fp$ models are compared to experiments in
Fig.~\ref{fig:uo2_spectra}.  The gap between the PES and IPES peaks is largely
determined by $U^{ff}$ (given the classification of UO$_2$ as a Mott-Hubbard
insulator) and is therefore sensitive to the value of $F^0$. Clearly, the
calculations both within the $f$-$f$ and $f$-$fp$ Hamiltonians severely
underestimate the gap. The SF model $F^0$ matches the experimental estimate of
$4.6 \pm 0.8$ eV to numerical precision \cite{baer_electronic_1980} and yields
spectra in excellent agreement with experiments \footnote{The variation in the
experimental spectra is likely due to differences in determination of the
chemical potential.  To match our cluster spectra intensity, the BIS spectra by
Baer and Schoenes \cite{baer_electronic_1980} were scaled 2.5x, close to the
factor of 2 suggested in their paper based on normalization considerations.}.

In summary, we have presented a scheme for a first-principles determination of
the effective Coulomb interactions for materials where
ligand-to-correlated-shell hybridizations and interactions cannot be neglected.
In the simplest version, intrashell interactions are reduced by the intershell
ones, and this estimate provides a remarkably good description of our test
materials.  Our scheme cures ambiguities in current cRPA calculations in the
case of entangled band structures.  Finally, our work builds a connection
between cluster model calculations, highly successful for spectroscopies of
strongly-correlated late TM oxides, and techniques such as DMFT and its
extensions treating the solid by embedding an effective local problem into an
auxiliary bath.

%%%%%%%%%%%%%%%%%%%%%%%%%%%%%%%%%%%%%%%%%%%%%%%%%%%%%%%%%%%%%%%%%%%%%%%%%%%%%%%%
We thank M. Haverkort for making his cluster calculation package
\cite{haverkort_multiplet_2012} available to us and useful discussions.  This
work was supported by the Chaire Energies Durables financed by EDF, by
IDRIS/GENCI under project 091393, by the French National Research Agency (ANR)
grant ANR-10-LABX-0039, and a Consolidator grant of the European Research
Council (project 617196).

%%%%%%%%%%%%%%%%%%%%%%%%%%%%%%%%%%%%%%%%%%%%%%%%%%%%%%%%%%%%%%%%%%%%%%%%%%%%%%%%
\bibliographystyle{plain}%merlin.mbs apsrev4-1.bst 2010-07-25 4.21a (PWD, AO, DPC) hacked

%%%%%%%%%%%%%%%%%%%%%%%%%%%%%%%%%%%%%%%%%%%%%%%%%%%%%%%%%%%%%%%%%%%%%%%%%%%%%%%%

%%%%%%%%%% Merge with supplemental materials %%%%%%%%%%
\widetext
\clearpage
\begin{center}
\textbf{\large Towards a first-principles determination of effective Coulomb
interactions in correlated electron materials: Role of intershell interactions
-- Supplementary Material}
\end{center}
\twocolumngrid
%%%%%%%%%% Merge with supplemental materials %%%%%%%%%%
%%%%%%%%%% Prefix a "S" to all equations, figures, tables and reset the counter %%%%%%%%%%
\setcounter{equation}{0}
\setcounter{figure}{0}
\setcounter{table}{0}
\setcounter{page}{1}
\makeatletter
\renewcommand{\theequation}{S\arabic{equation}}
\renewcommand{\thefigure}{S\arabic{figure}}
\renewcommand{\bibnumfmt}[1]{[S#1]}
\renewcommand{\citenumfont}[1]{S#1}
%%%%%%%%%% Prefix a "S" to all equations, figures, tables and reset the counter %%%%%%%%%%

In this supplementary material, we provide a brief review of the
constrained random phase approximation, additional background on our 
benchmark materials and further details on the computations presented
in the main part of the paper. We end with a discussion relating our
findings to current electronic structure approaches.

%%%%%%%%%%%%%%%%%%%%%%%%%%%%%%%%%%%%%%%%%%%%%%%%%%%%%%%%%%%%%%%%%%%%%%%%%%%%%%%%
%%%%%%%%%%%%%%%%%%%%%%%%%%%%%%%%%%%%%%%%%%%%%%%%%%%%%%%%%%%%%%%%%%%%%%%%%%%%%%%%
\section{Constrained random phase approximation}

The constrained random phase approximation as proposed in
\cite{aryasetiawan_frequency-dependent_2004s} has allowed for the first
principles determination of the effective local Coulomb interaction, the
``Hubbard $U$'', that is used as the bare interaction in a low-energy
Hamiltonian for a wide range of correlated materials ranging from transition
metals \cite{aryasetiawan_frequency-dependent_2004s,miyake_abinitio_2009s,
sasioglu_effective_2011s,shih_screened_2012s},
their oxides
\cite{miyake_screened_2008s,tomczak_realistic_2010s,martins_reduced_2011s,
vaugier_hubbard_2012s,sakuma_first-principles_2013s,werner_dynamical_2015s},
pnictides and chalcogenides
\cite{werner_satellites_2012s,nakamura_ab_2008s,miyake_d-_2008s,
imada_electronic_2010s,van_roekeghem_dynamical_2014s,van_roekeghem_inprep},
$f$-electron materials
\cite{karlsson_method_2010s,tomczak_rare-earth_2013s,nilsson_ab_2013s} to organic
conductors \cite{nomura_ab_2012s,nakamura_ab_2012s} and solid hydrogen
\cite{tomczak_effective_2009s}.  Probably the most important advantage of cRPA
over previous methods such as the constrained LDA
\cite{dederichs_ground_1984s,gunnarsson_density-functional_1989s,
anisimov_density-functional_1991s} or linear response
\cite{cococcioni_linear_2005s} schemes is the possibility of adapting $U$ to the
chosen low-energy model.  For reviews, see
\cite{biermann_dynamical_2014s,jiang_first-principles_2015s,imada_electronic_2010s}.

The underlying idea of the constrained random phase approximation is the
observation that the Hubbard $U$ should be obtained as the matrix element of a
{\it partially-screened} interaction in some localized basis.  This
partially-screened interaction is constructed such that screening at the RPA
level calculated for the low-energy Hamiltonian or for the full Coulomb
Hamiltonian lead to the same result for the fully-screened Coulomb interaction.

Mathematically, this can be obtained by constructing a partial polarization
function where certain screening processes are suppressed: $P_r = P - P_d$,
where $P_d$ is the polarization within the low-energy space, which at the RPA
level is a sum over particle-hole transitions that take entirely place within
the low-energy subspace. The partially-screened interaction is then obtained as
$W_r = v/(1-P_r v)$ where $v$ is the bare Coulomb interaction. Finally, the
matrix of Hubbard interactions $U_{m m^{'}m^{''} m^{'''}}$ is obtained as
matrix elements of $W_r$ within the basis of localized Wannier functions used
for the construction of the Hamiltonian.

Subtleties arise when one aims at constructing a multi-orbital Hamiltonian
which comprises more orbital degrees of freedom at the one-particle level than
in its interacting part.  The commonly applied strategy \cite{miyake_d-_2008s}
in this case is to construct Wannier functions for all orbitals, but exclude
screening transitions corresponding to the interacting degrees of freedom only.
However, ambiguities arise in this case when the correlated bands are entangled
with itinerant ones.  Several disentanglement strategies have been proposed
\cite{miyake_abinitio_2009s,nilsson_ab_2013s}, but it may appear somewhat
insatisfactory to deal with effectively an altered one-particle band structure,
in particular since by construction, the stronger the entanglement, the larger
the deviations.  Alternatively, one may think that in these cases it could be
advisable to rather reconsider the philosophy of cutting out only the few
correlated states out of a bigger number of entangled states living roughly at
the same energy.  Our work presents a well-defined solution following this
philosophy.

%%%%%%%%%%%%%%%%%%%%%%%%%%%%%%%%%%%%%%%%%%%%%%%%%%%%%%%%%%%%%%%%%%%%%%%%%%%%%%%%
%%%%%%%%%%%%%%%%%%%%%%%%%%%%%%%%%%%%%%%%%%%%%%%%%%%%%%%%%%%%%%%%%%%%%%%%%%%%%%%%
\section{Background on the benchmark compounds}

%%%%%%%%%%%%%%%%%%%%%%%%%%%%%%%%%%%%%%%%%%%%%%%%%%%%%%%%%%%%%%%%%%%%%%%%%%%%%%%%
\subsection{Cerium}

Elemental cerium has intrigued the solid state community for decades due to its
isostructural volume-collapse transition.  This ``$\alpha$-$\gamma$''
transition is of first order below a critical temperature T$_c$ = 600K, and
ends in a second-order critical point at T$_c$.  At room temperature the
decrease in volume when going from the $\gamma$ to the $\alpha$ phase is as
large as 15$\%$. A Curie-Weiss magnetic susceptibility in the $\gamma$ phase
signals the presence of localized $4f$ electrons that delocalize in the
$\alpha$ phase where the magnetic susceptibility becomes Pauli-like.  In this
phase, the $4f$ electrons participate in the bonding (resulting in the smaller
volume) and the formation of quasiparticles, as seen in photoemission
\cite{wuilloud_electronic_1983s,wieliczka_high-resolution_1984s}.

The origin of this transition has been the subject of intense debate in the
literature. The ``Mott transition picture'' interprets it as a
delocalization-localization transition in the $4f$ shell only
\cite{johansson_a-g_1974s}.  The ``Kondo volume collapse'' (KVC) picture
\cite{allen_kondo_1982s,allen_alpha-gamma_1992s,lavagna_volume_1982s} in contrast
attributes the leading role to the hybridization between the $4f$ electrons and
($spd$-) itinerant states: the stronger hybridization in the $\alpha$-phase
leads to a high Kondo temperature and thus to screening of the $4f$ local
moment, while the high-volume $\gamma$-phase has a low Kondo temperature,
leading in practice to unscreened moments at intermediate temperatures.

Cerium has long served as a test material for methods that go beyond the LDA
\cite{sakuma_self-energy_2012s,casadei_density-functional_2012s} and in
particular for DMFT-based electronic structure calculations
\cite{held_cerium_2001s,zolfl_spectral_2001s,mcmahan_thermodynamic_2003s,
haule_cerium_2005s,amadon_cerium_2006s,streltsov_magnetic_2012s,litsarev_correlated_2012s,
bieder_thermodynamics_2014s,lanata_interplay_2014s,chakrabarti_alpha-gamma_2014s}.
Combined DFT+DMFT was found to be able to disentangle subtle
electronic-structure changes in the hybridization to elucidate the mechanism of
the transition, and give a description of spectral, optical and magnetic
properties in agreement with experiments.  It was also emphasized, however,
that at any relevant temperature the entropic contribution dominates over the
internal energy changes and pressure contribution such that the question of the
existence of a double minimum in the internal energy acquires a somewhat
academic character.  We mention very recent QMC calculations
\cite{devaux_electronic_2015s} that advocate the existence of a transition down
to zero temperature. 

With the advent of more efficient Monte Carlo techniques for the DMFT equations
in the multi-orbital case in recent years, the problem has been extensively
revisited, and not only total energies, but also entropy and free energy have
been computed \cite{bieder_thermodynamics_2014s}.  Recent work has also focused
on the role of spin-orbit coupling
\cite{dong_orbital-dependent_2014s,lanata_interplay_2014s}.

While many-body calculations for cerium have thus been extremely successful,
$U$ values have generally been used as adjustable parameters, and several
authors have discussed the difficulties in determining the effective Hubbard
interactions from first principles.  This comes without surprise since this
material can be considered a ``worst-case'' scenario, precisely due to the
strong hybridization of the $4f$-states with itinerant conduction band states
present at the same energy.  Ref.~\onlinecite{amadon_screened_2014s} has
presented a detailed study of various choices of low-energy models, arriving at
the conclusion that any low-energy Hamiltonian of $f$-$spdf$ type would have
$U^{ff}$-values lower than 1 eV in cRPA. This illustrates the problem: there is
a clear consensus in the literature that a much larger value of 5 to 6 eV is
necessary to reproduce the available experimental findings.  Nilsson et
al.~\cite{nilsson_ab_2013s} proposed a disentanglement scheme which, by
essentially replacing the original band structure by an approximate one with
reduced hybridization, yields larger (though still too small) values of 4.3 eV
and 5.4 eV for $\alpha$ and $\gamma$ cerium respectively.

%%%%%%%%%%%%%%%%%%%%%%%%%%%%%%%%%%%%%%%%%%%%%%%%%%%%%%%%%%%%%%%%%%%%%%%%%%%%%%%%
\subsection{NiO}

NiO has been put forward as a challenge to solid state theory and electronic
structure calculations since the very early days: Mott highlighted the
inadequacies of the band picture in describing the physics of this compound.
NiO is an antiferromagnetic insulator below a N\'eel temperature of 525 K.
Above T$_N$, the antiferromagnetic order disappears, but the material remains
insulating with a large gap of 4 eV that largely exceeds the energy scale of
the N\'eel temperature). Furthermore, its spectral properties are remarkably
insensitive to the presence or absence of magnetic order. This calls for a
theoretical description of the insulating nature without relying on magnetic
order.

The story became more involved with the seminal work of Zaanen, Sawatzky and
Allen \cite{zaanen_band_1985s}, who distinguished between Mott insulators where
the gap opens due to the Coulomb blocking between correlated states of the same
shell character, and charge-transfer insulators where the Hubbard interactions
are larger than the O $2p$ to Ni $3d$ charge-transfer energy such that the
first ionization states are in fact of O $p$ character.

In NiO, it was already known that a satellite feature at -8 eV binding energy
is mainly of $d$ character, while the low-energy states at the Fermi level have
large oxygen $2p$ character.  A series of photoemission (including resonant
photoemission) works elucidated the different transitions in detail
\cite{hufner_photoemission_1984s,sawatzky_magnitude_1984s,fender_covalency_1968s,
cheetham_magnetic_1983s,shen_aspects_1990s,shen_electronic_1991s,shen_angle_1991s,
fujimori_multielectron_1984s,tjeng_giant_1991s,fink_electron_1994s}.

NiO was treated within many-body perturbation theory within the GW
approximation by Aryasetiawan and Gunnarsson
\cite{aryasetiawan_electronic_1995s}.  The authors found the gap to open
within this approximation, but also pointed out failures concerning the gap
character which remains Mott-Hubbard like, and satellite structures related to
the strongly-atomic character of the $3d$ states.

The first DMFT works on NiO used a Hamiltonian where only the TM $3d$ states
were included \cite{ren_lda+dmft_2006s}. Soon thereafter, it was pointed out
that this was questionable due to the $d$-ligand interplay at the Fermi level.
Several subsequent works included ligand states as itinerant states in the
one-body Hamiltonian \cite{korotin_construction_2008s,kunes_nio:_2007s}.
However, none of the standard double counting corrections appeared to be
appropriate in this case \cite{karolak_double_2010s}, and NiO still appears to
be a challenge to DMFT-based or DMFT-like theories
\cite{manghi_multi-orbital_2014s}.

%%%%%%%%%%%%%%%%%%%%%%%%%%%%%%%%%%%%%%%%%%%%%%%%%%%%%%%%%%%%%%%%%%%%%%%%%%%%%%%%
\subsection{Actinide oxides: UO$_2$, PuO$_2$ and NpO$_2$}

The actinide dioxides UO$_2$, PuO$_2$ and NpO$_2$ are insulating materials with
similar gaps (of 
2.1 eV \cite{schoenes_optical_1978s},
2.9 eV \cite{mccleskey_optical_2013s}
and 
2.8 eV \cite{mccleskey_optical_2013s}
respectively, but quite different ground states.  UO2 has been and remains the
subject of intense interest from the applied materials perspective thanks to
its application as a nuclear fuel. Electronic structure calculations thus focus
not only on questions of stability of different structures, but also attempt to
assess the physics of impurities resulting from the radioactive decay
processes.  We mention for example the study of He atom incorporation into
UO$_2$ crystals by \cite{gryaznov_ab_2009s}, where LDA+U with $U=4.6$ eV was
employed.  For an extensive review and recent work see
\cite{vathonne_etude_2014s}.  Interestingly, corrections to simple DFT-LDA
calculations turn out to be necessary not only for the description of spectral
properties but even in assessing the energetics, and UO$_2$ has become a
playground for electronic structure approaches.  While the value of $U$ has
typically been considered an adjustable parameter or guessed from spectroscopy
experiments, we note that the values employed coincide remarkably well with our
results discussed in the main text.

The series of actinide dioxides has also attracted much interest due to their
exotic low-temperature (multi-polar) ordered states.  While in PuO$_2$ the
$f^4$ configuration leads to a non-magnetic ground state, UO$_2$ has a complex
antiferromagnetic order below a N\'eel temperature of 30.8 K
\cite{wen_density_2013s}, with a transverse 3-k arrangement driven by
quadrupolar interactions. NpO$_2$ on the other hand displays a longitudinal 3-k
order below an ordering temperature of 25 K, with a rank-5 magnetic multipole
(``triakontadipole'').  Despite being difficult due to the noncollinear
character of the order and high-rank order parameters, LDA+U calculations are
successful in describing the multi-polar order with interaction values of $U=4$
eV \cite{suzuki_first-principles_2010s}.

Their complex electronic structure has always been considered a challenge also
for purely functional-based first-principles calculations, and it is not
surprising that the actinide oxides became a playground and testbed for hybrid
functionals (see e.g., \cite{kudin_hybrid_2002s,roy_dispersion_2008s}).  A
comprehensive review of various electronic structure calculations of the
actinide oxides is given in \cite{wen_density_2013s}.

%%%%%%%%%%%%%%%%%%%%%%%%%%%%%%%%%%%%%%%%%%%%%%%%%%%%%%%%%%%%%%%%%%%%%%%%%%%%%%%%
%%%%%%%%%%%%%%%%%%%%%%%%%%%%%%%%%%%%%%%%%%%%%%%%%%%%%%%%%%%%%%%%%%%%%%%%%%%%%%%%
\section{Detailed derivation of the shell-folded Hamiltonian}

Here we present the derivation of the shell-folded Hamiltonian by the
Peierls-Feynman-Bogoliubov variational principle in a step-by-step manner and
determine the form of our effective interactions 
$\tilde{U}^{dd} = U^{dd} - U^{pd}$ and $\tilde{U}^{pp} = U^{pp} - U^{pd}$. 

Consider the $dp$ Hamiltonian
\begin{align*} 
 h_{\bf R} = & \frac{\tilde{U}^{dd}}{2} \sum_{\begin{subarray}{c} 
                                                       (m\s) \ne (m'\s^{\prime})\\
                                                       m,m' \in {\cal C}
                                                      \end{subarray}}
                                            \hspace{-.4cm} 
n_{m \s} n_{m' \s'} 
+ \frac{\tilde{U}^{pp}}{2} \sum_{\begin{subarray}{c} 
                                                       (m\s) \ne (m'\s^{\prime})\\
                                                       m,m' \in {\cal L}
                                         \end{subarray}}
                                            \hspace{-.4cm} 
                                                      n_{m \s} n_{m' \s'}
\\  + & 
  \tilde{U}^{dp}           \sum_{\begin{subarray}{c} 
                                                       m \in {\cal C}, \s \\
                                                       m' \in {\cal L}, \s^{\prime}
                                         \end{subarray}}
                                            \hspace{-.4cm} 
                                                      n_{m \s} n_{m' \s'},
\end{align*}

and the auxiliary Hamiltonian
\begin{align*} 
 \tilde{h}_{\bf R} = & \frac{\tilde{U}^{dd}}{2} \sum_{\begin{subarray}{c} 
                                                       (m\s) \ne (m'\s^{\prime})\\
                                                       m,m' \in {\cal C}
                                                      \end{subarray}}
                                            \hspace{-.4cm} 
n_{m \s} n_{m' \s'} 
+ \frac{\tilde{U}^{pp}}{2} \sum_{\begin{subarray}{c} 
                                                       (m\s) \ne (m'\s^{\prime})\\
                                                       m,m' \in {\cal L}
                                         \end{subarray}}
                                            \hspace{-.4cm} 
                                                      n_{m \s} n_{m' \s'}.
\end{align*}
where $\tilde{U}^{dd/pp}$ are the effective Coulomb interactions for the $d$
and $p$ manifolds. We find their optimal values by minimizing $\tilde{F} =
F[\tilde{h}_{\bf R}] + \ave{h_{\bf R} - \tilde{h}_{\bf R}}_0$ with respect to
$\tilde{U}^{dd}$ and $\tilde{U}^{pp}$, where $\ave{\ldots}_0$ refers to the
average taken with the auxiliary Hamiltonian. 
Let us also extend the definition
\begin{eqnarray*}
D_Z^{XY} = \frac{\partial}{\partial \tilde{U}^{Z}}
         \sum_{(i\s) \neq (j\s')} \ave{ n^{X}_{i \s} n^{Y}_{j \s'}}_0 .
\end{eqnarray*}
where $X,Y,Z=d$ or $p$.

Then, from the stationarity condition
$ \partial \tilde{F} / \partial \tilde{U}^{dd} = 0 $
we obtain
\begin{align*}
\label{eq:U0_shellfolded}
\tilde{U}^{dd} = U^{dd} + 
\frac{ 2 U^{dp} D_d^{dp}
       + (U^{pp} - \tilde{U}^{pp}) D_d^{pp}
      }{D_d^{dd}}.
\end{align*}
and an analogous equation for $\tilde{U}^{pp}$.

Solving the system of equations, we have for 
$\Delta = \tilde{U}^{dd} - U^{dd}$,
\begin{align*}
\Delta & = 
\frac{\Delta D_p^{dd} - 2 U^{dp} D_p^{dp}}
     {D_p^{pp}}
\frac{D_d^{pp}}
     {D_d^{dd}}
+
\frac{2 U^{dp} D_d^{dp} }
     {D_d^{dd}} 
\end{align*}

Simplifying this expression gives
\begin{align*}
\frac{\tilde{U}^{dd} - U^{dd}}
     {U^{pd}} & = 
\frac{2 D_d^{pd} D_p^{pp} - 2 D_p^{pd} D_d^{pp}}
     {  D_d^{dd} D_p^{pp} -   D_p^{dd} D_d^{pp}}.
\end{align*}

Assuming that the total number of particles is invariant with respect to
changes in the effective parameters, namely that
\begin{align*}
  \frac{\partial \ave{N(N-1)}}{\partial \tilde{U}^{dd}} 
= \frac{\partial \ave{N(N-1)}}{\partial \tilde{U}^{pp}} 
= 0,
\end{align*}
or equivalently that
\begin{align*}
 D_d^{dd} + 2 D_d^{dp} + D_d^{pp} = 0,\\
 D_p^{dd} + 2 D_p^{dp} + D_p^{pp} = 0,
\end{align*}
we can simply the above to
\begin{align*}
\frac{\tilde{U}^{dd} - U^{dd}} {U^{pd}} & = -1.
\end{align*}

Thus, we obtain that
\begin{align*}
\tilde{U}^{dd} = U^{dd} - U^{pd}, \\
\tilde{U}^{pp} = U^{pp} - U^{pd}.
\end{align*}

While the above derivation is presented for spin- and orbital-averaged
interactions for simplicity, it is equally applicable to the orbital-resolved
interactions $U_{m \s m' \s'}$. As long as we assume the intershell interaction
to be an orbital-independent average $U^{dp}$, the resulting effective
interactions are obtained from an effective reduced Slater integral
$\tilde{F}^0 = F^0 - U^{dp}$.  Note that quadratic terms $H_0$ are unchanged.
As we are devising an effective model for the \emph{local} $h_{\bf R}$, and as
our summations run explicitly over all orbitals contained therein, we do not
have to consider either the number of equivalent atoms within the unit cell or
its coordination.

%%%%%%%%%%%%%%%%%%%%%%%%%%%%%%%%%%%%%%%%%%%%%%%%%%%%%%%%%%%%%%%%%%%%%%%%%%%%%%%%
%%%%%%%%%%%%%%%%%%%%%%%%%%%%%%%%%%%%%%%%%%%%%%%%%%%%%%%%%%%%%%%%%%%%%%%%%%%%%%%%
\section{Construction of transition metal centred Wannier functions}

In the principal part of the Letter, we derived a low-energy model that we
showed to be successful in treating intershell interactions in addition to the
intrashell interactions.  Here, using NiO as an example, we describe the
construction of the ``ring'' orbitals centred on the correlated site as a
superposition of the ligand $p$ orbitals.  This ring orbital construction is
inspired by the cell-perturbation methods
\cite{jefferson_derivation_1992s,feiner_effective_1996s,raimondi_effective_1996s}
that emerged as a consequence of the Zhang-Rice construction
\cite{zhang_effective_1988s} for the Cu superconductors.

%%%%%%%%%%%%%%%%%%%%%%%%%%%%%%%%%%%%%%%%%%%%%%%%%%%%%%%%%%%%%%%%%%%%%%%%%%%%%%%%
%%%%%%%%%%%%%%%%%%%%%%%%%%%%%%%%%%%%%%%%%%%%%%%%%%%%%%%%%%%%%%%%%%%%%%%%%%%%%%%%
\section{NiO}

%%%%%%%%%%%%%%%%%%%%%%%%%%%%%%%%%%%%%%%%%%%%%%%%%%%%%%%%%%%%%%%%%%%%%%%%%%%%%%%%
\subsection{Construction of ring orbitals}

NiO adopts the rock-salt structure with a face-centred cubic lattice of Ni atoms 
intercalated with a face-centred cubic lattice of O atoms. As the conventional unit 
cell contains four formula units, for simplicity, we choose to work globally in
the Cartesian coordinate system with the primitive unit cell vectors
\begin{align*}
 {\bf a} = (1/2, 1/2, 0), \\ 
 {\bf b} = (1/2, 0, 1/2), \\
 {\bf c} = (0, 1/2, 1/2).
\end{align*}
Our basis consists of one Ni atom at $(0,0,0)$ and one O atom at $(1/2, 1/2,
1/2)$.  We define the corresponding $k$-vectors ${\bf k_{a}}, {\bf k_{b}}, {\bf
k_{c}}$ such that ${\bf k_{i}} \cdot {\bf j} = 2\pi \delta_{ij}$.

In order to find matrix elements of $H(\mathbf{k})$, we first determine the Ni
$3d$ to O $2p$ hopping directions. Using the Slater-Koster parametrization
\cite{slater_simplified_1954s}, we write the matrix elements of the real-space
tight-binding Hamiltonian $H(\mathbf{R})$ as functions of the $p$-$p$, $p$-$d$
and $d$-$d$ two-center bond integrals that give the strength of bonding in a
given direction, reflecting the relative orbital lobe positions. We use the
standard notation of $\sigma$-bonding if the lobes point to each other and
$\pi$-bonding if the lobes are parallel. The resulting matrix is finally
Fourier-transformed to $H(\mathbf{k})$ in momentum space. We note that, besides
hopping, the cubic crystal field splitting of Ni $3d$-states into $\eg$ and
$\ttg$ (parametrized in the usual notation as $10Dq$) is also included in
$H(\mathbf{k})$. The final $H(\mathbf{k})$ in matrix form is given in Table
\ref{tab:hk}. The diagonalization of this Hamiltonian yields the tight-binding
band structure.

\begin{sidewaystable}[htdp] \scriptsize
\caption{
The Hamiltonian operator in momentum space, $H(\mathbf{k})$, in the basis
$(\text{O} \, 2p_x, \text{O} \, 2p_y, \text{O} \, 2p_z,
\text{Ni} \, 3d_{xy}, \text{Ni} \, 3d_{yz}, \text{Ni} \, 3d_{xz}, 
\text{Ni} \, 3d_{x^2-y^2}, \text{Ni} \, 3d_{z^2})$. 
To improve readability we employ the abbreviations 
$s_i = \sin(\frac{k_i}{2})$ and $c_i = \cos(\frac{k_i}{2})$.
}
\label{tab:hk}
\centering
\begin{tabular}{cccccccc}
 $ \epsilon_p+2 \text{pp$\sigma $} c_x \left(c_y+c_z\right) $ &
 $ -2 \text{pp$\sigma $} s_x s_y $ &
 $ -2 \text{pp$\sigma $} s_x s_z $ &
 $ -2 i \text{pd$\pi $} s_y $ &
 $ 0 $ &
 $ -2 i \text{pd$\pi $} s_z $ &
 $ -i \sqrt{3} \text{pd$\sigma $} s_x $ &
 $ i \text{pd$\sigma $} s_x $ \\
 $ -2 \text{pp$\sigma $} s_x s_y $ &
 $ \epsilon_p+2 \text{pp$\sigma $} (c_x c_y + c_y c_z) $ &
 $ -2 \text{pp$\sigma $} s_y s_z $ &
 $ -2 i \text{pd$\pi $} s_x $ &
 $ -2 i \text{pd$\pi $} s_z $ &
 $ 0 $ &
 $ i \sqrt{3} \text{pd$\sigma $} s_y $ &
 $ i \text{pd$\sigma $} s_y $ \\
 $ -2 \text{pp$\sigma $} s_x s_z $ &
 $ -2 \text{pp$\sigma $} s_y s_z $ &
 $ \epsilon_p+2 \text{pp$\sigma $} (c_x c_z + c_y c_z) $ &
 $ 0 $ &
 $ -2 i \text{pd$\pi $} s_y $ &
 $ -2 i \text{pd$\pi $} s_x $ &
 $ 0 $ &
 $ -2 i \text{pd$\sigma $} s_z $ \\
 $ 2 i \text{pd$\pi $} s_y $ &
 $ 2 i \text{pd$\pi $} s_x $ &
 $ 0 $ &
 $ -4 \text{Dq}+\epsilon_d+3 \text{dd$\sigma $} c_x c_y $ &
 $ 0. $ &
 $ 0. $ &
 $ 0 $ &
 $ \sqrt{3} \text{dd$\sigma $} s_x s_y $ \\
 $ 0 $ &
 $ 2 i \text{pd$\pi $} s_z $ &
 $ 2 i \text{pd$\pi $} s_y $ &
 $ 0. $ &
 $ -4 \text{Dq}+\epsilon_d+3 \text{dd$\sigma $} c_y c_z $ &
 $ 0. $ &
 $ \frac{3}{2} \text{dd$\sigma $} s_y s_z $ &
 $ -\frac{\sqrt{3}}{2} \text{dd$\sigma $} s_y s_z $ \\
 $ 2 i \text{pd$\pi $} s_z $ &
 $ 0 $ &
 $ 2 i \text{pd$\pi $} s_x $ &
 $ 0. $ &
 $ 0. $ &
 $ -4 \text{Dq}+\epsilon_d+3 \text{dd$\sigma $} c_x c_z $ &
 $ -\frac{3}{2} \text{dd$\sigma $} s_x s_z $ &
 $ -\frac{\sqrt{3}}{2} \text{dd$\sigma $} s_x s_z $ \\
 $ i \sqrt{3} \text{pd$\sigma $} s_x $ &
 $ -i \sqrt{3} \text{pd$\sigma $} s_y $ &
 $ 0 $ &
 $ 0 $ &
 $ \frac{3}{2} \text{dd$\sigma $} s_y s_z $ &
 $ -\frac{3}{2} \text{dd$\sigma $} s_x s_z $ &
 $ 6 \text{Dq}+\epsilon_d+\frac{3}{4} \text{dd$\sigma $} (c_x c_z + c_y c_z) $ &
 $ \frac{\sqrt{3}}{4} \text{dd$\sigma $} \left(c_x-c_y\right) c_z $ \\
 $ -i \text{pd$\sigma $} s_x $ &
 $ -i \text{pd$\sigma $} s_y $ &
 $ 2 i \text{pd$\sigma $} s_z $ &
 $ \sqrt{3} \text{dd$\sigma $} s_x s_y $ &
 $ -\frac{\sqrt{3}}{2} \text{dd$\sigma $} s_y s_z $ &
 $ -\frac{\sqrt{3}}{2} \text{dd$\sigma $} s_x s_z $ &
 $ \frac{\sqrt{3}}{4} \text{dd$\sigma $} \left(c_x-c_y\right) c_z $ &
 $ 6 \text{Dq}+\epsilon_d+\frac{1}{4} \text{dd$\sigma $} (c_y c_z + c_x \left(4 c_y+c_z\right)) $ \\
\end{tabular}
\end{sidewaystable}

\begin{figure}[htpb]
  \centering
    \includegraphics[width=1.00\columnwidth]{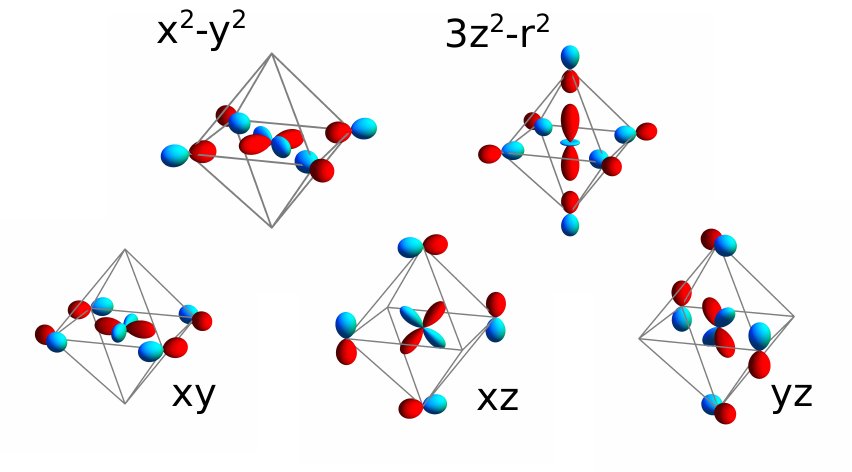}
    \caption{The NiO$_6$ octahedron cluster orbitals.
             \label{fig:nio_cluster_orb}}
\end{figure}

We now construct Ni-centred ring orbitals from the three O $2p$ states. This is
similar to what is done in typical cluster calculation ligand orbitals
\cite{haverkort_multiplet_2012s}. The bonding cluster orbitals corresponding to the 
NiO$_6$ octahedron are shown in Fig.~\ref{fig:nio_cluster_orb}.

\begin{figure}[htpb]
  \centering
    \includegraphics[width=1.00\columnwidth]{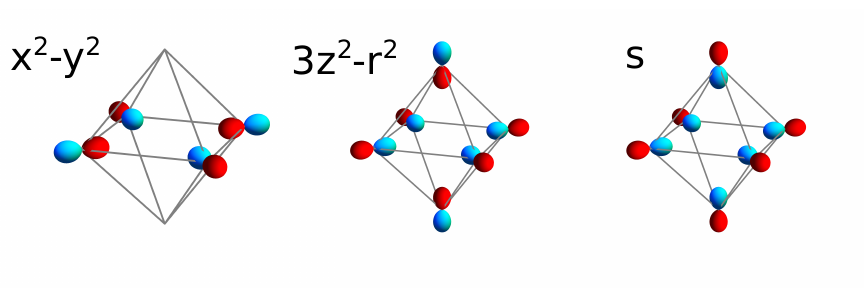}
    \caption{The NiO ring orbitals. The first two orbitals have $\eg$ symmetry
             while the last is of $a_{1g}$ symmetry.
             \label{fig:nio_ring_orb}}
\end{figure}

One key difference between our ring orbitals and the cluster orbitals, however,
is that the cluster orbitals are constructed entirely in real space from
combinations of a finite number of orbitals. In our case, we use a coherent
superposition of orbitals on an all (infinite) sites, giving rise to a
$k$-dependence (see the $\sin$ terms in Eq.~\ref{eq:p_to_ring}).  From the
three available oxygen states we can construct two $\eg$-type rings, which
locally hybridize with the Ni $\eg$ orbitals and a third, $a_{1g}$ (i.e.,
$s$-like) ring which is locally non-bonding with the Ni $\eg$ orbitals.  The
form of these rings is shown in Fig.~\ref{fig:nio_ring_orb}.

Mathematically, we perform a unitary transformation on the $2p$ subspace of the
original Hamiltonian $H$ to $\tilde{H}$ by $\tilde{H} = T H {T}^\dagger$. The
transformation matrix $T$ takes us from the
$(p_x,p_y,p_z,d_{xy},d_{yz},d_{xz},d_{x^2-y^2},d_{z^2})$ to
$(\tilde{p}_{x^2-y^2},\tilde{p}_{z^2},\tilde{p}_s,d_{xy},d_{yz},d_{xz},d_{x^2-y^2},d_{z^2})$
basis, and is given by 

\begin{equation}
\label{eq:p_to_ring}
T = \begin{pmatrix}
  \frac{s_x}{\sqrt{2}}  & 
 -\frac{s_y}{\sqrt{2}}  & 
 0 & 0 & 0 & 0 & 0 & 0 \\
  \frac{s_y}{\sqrt{6}}  & 
  \frac{s_x}{\sqrt{6}}  & 
  \frac{-2 s_z}{\sqrt{6}}  & 
 0 & 0 & 0 & 0 & 0 \\
  \frac{s_y}{\sqrt{3}}  & 
  \frac{s_x}{\sqrt{3}}  & 
  \frac{s_z}{\sqrt{3}}  & 
 0 & 0 & 0 & 0 & 0 \\
 0 & 0 & 0 & 1 & 0 & 0 & 0 & 0 \\
 0 & 0 & 0 & 0 & 1 & 0 & 0 & 0 \\
 0 & 0 & 0 & 0 & 0 & 1 & 0 & 0 \\
 0 & 0 & 0 & 0 & 0 & 0 & 1 & 0 \\
 0 & 0 & 0 & 0 & 0 & 0 & 0 & 1 \\
\end{pmatrix}
\end{equation}

We have used the abbreviation $s_i = \sin(\frac{k_i}{2})$.  The transformation
is chosen such that the ring with $d_{x^2-y^2}$ symmetry is exact, while the
$d_{z^2}$ and $s$-like rings are approximate in order to orthogonalize them.

\begin{figure}[htpb]
  \centering
    \includegraphics[width=1.00\columnwidth]{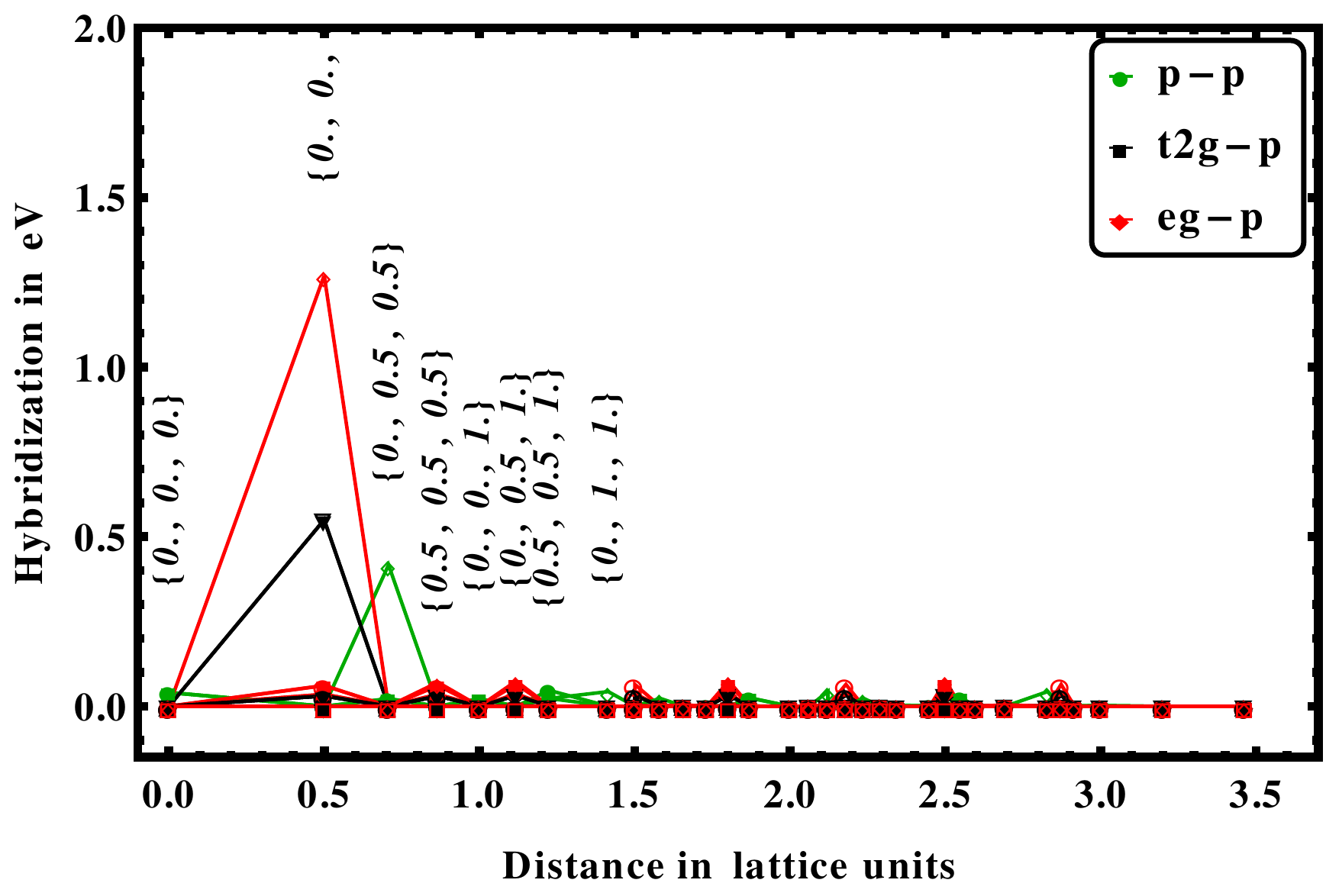}
    \caption{Matrix elements of the initial $dp$ Hamiltonian $H(\mathbf{k})$ in
             real space plotted as a function of distance (measured in conventional unit
             cell lengths).
             \label{fig:p_hopping}}
\end{figure}

In the original O-centred $p$ basis, the hybridization elements decay with
distance as shown in Fig.~\ref{fig:p_hopping}. Note that the distances are
measured in conventional lattice units. As we choose to construct a
maximally-symmetric primitive cell as described above, the O atoms closest to
the Ni atom at (0,0,0) actually lie in the nearest-neighbour unit-cells rather
than within the (0,0,0) unit cell. This is reflected in the peak in $t_{\ttg
p}$ and $t_{\eg p}$ at (0,0,$\frac{1}{2}$). As the $d$-subspace of the
Hamiltonian is untouched by the transformation, the $t_{d d}$ hybridizations in
the original $p$ basis and in the ring basis are identical.

\begin{figure}[htpb]
  \centering
    \includegraphics[width=1.00\columnwidth]{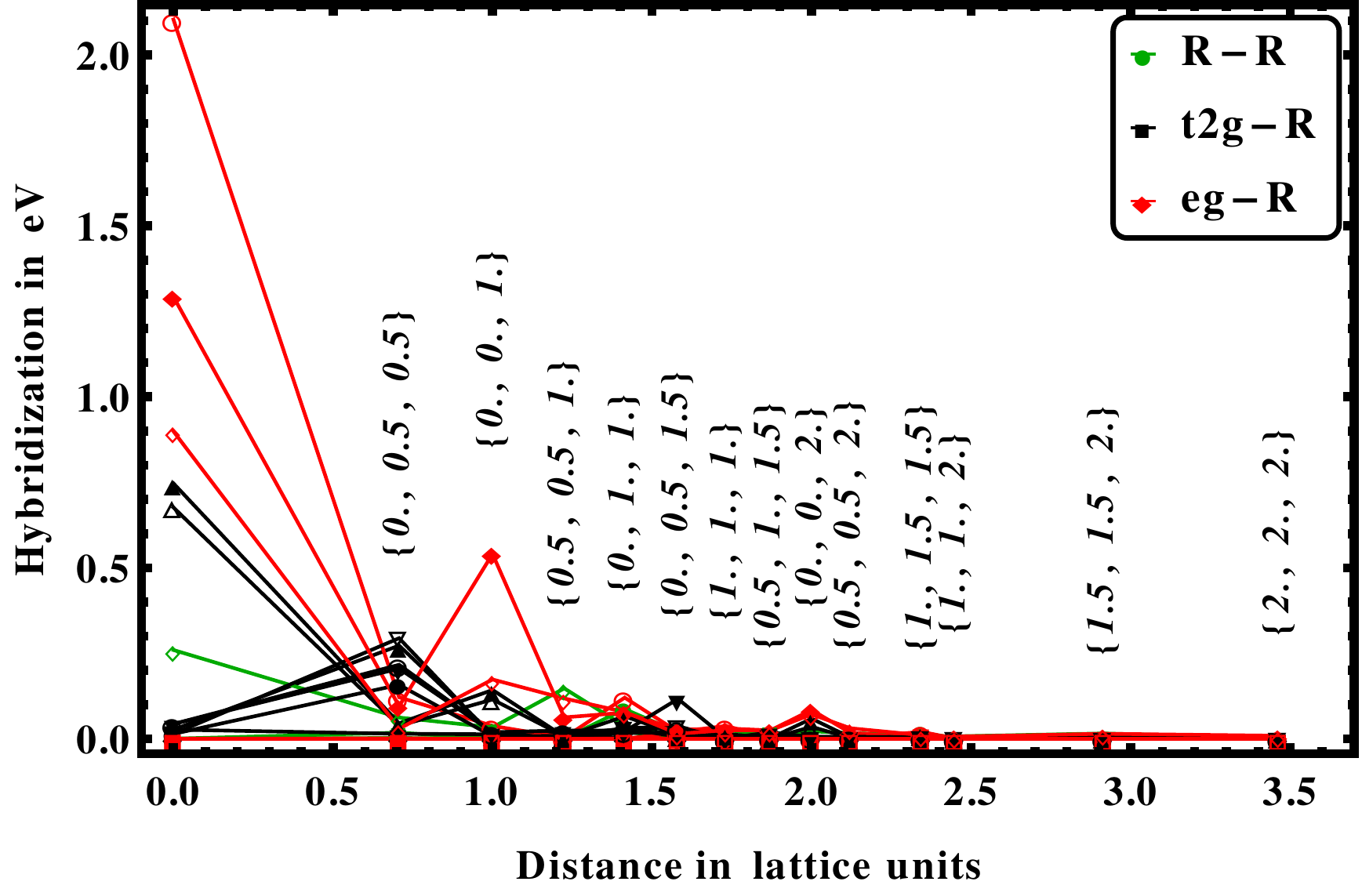}
    \caption{Matrix elements of the transformed $d$-ring Hamiltonian
             $\tilde{H}(\mathbf{k})$ in real space plotted as a function of distance
             (measured in conventional unit cell lengths).
      \label{fig:ring_hopping}}
\end{figure}

In the Ni-centred ring basis, the decay of the hybridization elements is
plotted in Fig.~\ref{fig:ring_hopping}. The $s$-like ring orbital is always
non-bonding locally; however, given that the symmetry of this orbital is not
imposed exactly in order to ensure unitarity of the transformation, it
continues to hybridize weakly with the Ni $d$ orbitals. For further symmetry
reasons, the $d_{x^2-y^2}$-like ring orbital alternates between being bonding
and anti-bonding over second nearest-neighbours.

The transformation of the $p$ orbitals into ring orbitals also modifies the
interaction parameters $U^{pp}$. Notably, the intra-ring interaction decreases
as the rings become more delocalized, while inter-ring interactions increase as
overlap between orbitals increases. The ring-$d$ interactions are unchanged due
to symmetry reasons.

%%%%%%%%%%%%%%%%%%%%%%%%%%%%%%%%%%%%%%%%%%%%%%%%%%%%%%%%%%%%%%%%%%%%%%%%%%%%%%%%
%%%%%%%%%%%%%%%%%%%%%%%%%%%%%%%%%%%%%%%%%%%%%%%%%%%%%%%%%%%%%%%%%%%%%%%%%%%%%%%%
\section{UO2}

\begin{figure}[htpb]
  \centering
    \includegraphics[width=1.00\columnwidth]{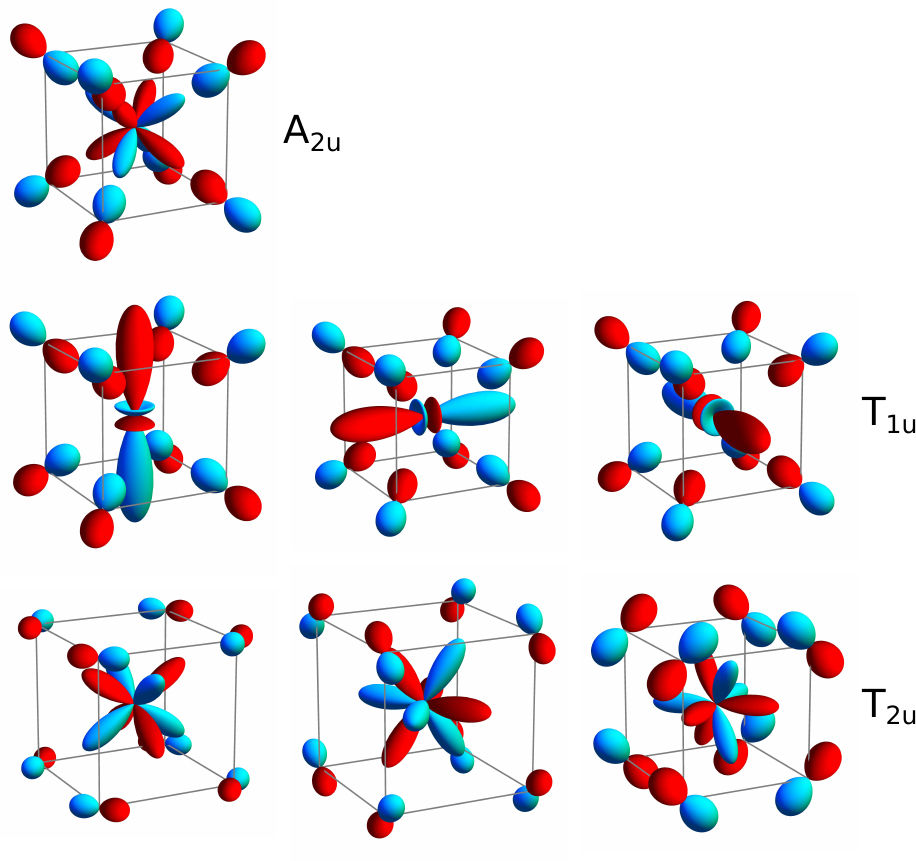}
    \caption{The 7 bonding ligand orbitals constructed as linear combinations
             of the 8 $p$ orbitals with the symmetry of the corresponding $f$ orbital.
             \label{fig:uo2_cluster_orb}}
\end{figure}

In Fig.~\ref{fig:uo2_cluster_orb}, we show the 7 bonding ligand orbitals for
the eight-fold coordinated U $5f$ states. We can construct ligand orbitals by
linear combination of O $2p$ states matching the respective symmetry of the U
$5f$ orbitals as was done for NiO above. However, unlike for a finite size
cluster, we cannot create Bloch states associated to each of the ligand
orbitals (in the unit cell we have only six oxygen states).  Hence, in order to
construct ring-orbitals one has to choose which six basis states to include, as
in NiO. 

Finally, we give the computational details for the cluster model calculations
for UO$_2$ that result in the spectral functions shown in Fig.~2 of the main
text. In the calculation, both $f$- and $p$-states, separated by the
charge-transfer energy given by LDA, are included but are not allowed to
hybridize. The parameters used for the UO$_2$ cluster model are 
$E_{a_u} = 0.349$, 
$E_{t_{1u}} = -0.039$, 
$E_{t_{2u}} = -0.077$, 
$\Delta_{fp} = 5.189$,
which were obtained from LDA and $\zeta_{5f} = 0.27$, obtained from an atomic
HF calculation. The hybridization parameters $V(a_u)=V(t_{1u})=V(t_{2u})=0$.
All values are in eV. 

%%%%%%%%%%%%%%%%%%%%%%%%%%%%%%%%%%%%%%%%%%%%%%%%%%%%%%%%%%%%%%%%%%%%%%%%%%%%%%%%
%%%%%%%%%%%%%%%%%%%%%%%%%%%%%%%%%%%%%%%%%%%%%%%%%%%%%%%%%%%%%%%%%%%%%%%%%%%%%%%%
\section{Overview of U values used in the literature}

\begin{table}[htdp]
\caption{Screened Coulomb interactions used in the literature. All values in eV.}
\label{tab:literature_u}
\centering
\begin{tabular}{ccc}
\multicolumn{2}{c}{Interaction parameters} & References \\
\hline
$U = 4.5$         &  $J = 0.51$        &   \onlinecite{dorado_dft+u_2009s,geng_structural_2007s,geng_first-principles_2011s} \\
$\bar{U} = 4.5$   &                    &   \onlinecite{dudarev_understanding_2000s} \\
$\bar{U} = 4.5$   &  $\bar{J} = 0.51$  &   \onlinecite{dudarev_electronic_1998s} \\
$U = 4.5$         &  $J = 0.54$        &   \onlinecite{dudarev_effect_1997s} \\
$U = 4.6$         &  $J = 0.5 $        &   \onlinecite{gryaznov_density_2010s} \\
$U = 7.7$         &                    &   \onlinecite{gunnarsson_electronic_1988s} \\
$U = 4.0$         &  $J = 0.51$        &   \onlinecite{hongliang_optical_2010s} \\
$U = 4.5$         &  $J = 0.5 $        &   \onlinecite{jollet_electronic_1997s} \\
$U = 3.0,6.5$     &  $J = 0.7 $        &   \onlinecite{kolorenc_electronic_2015s} \\
$U = 4.5-4.75$    &  $J = 0.54$        &   \onlinecite{kotani_systematic_1992s} \\
$U = 4.0$         &                    &   \onlinecite{petit_electronic_2010s} \\
$U = 8.0$         &                    &   \onlinecite{shim_x-ray_2009s} \\
$U = 6.0$         &  $J = 0.5-0.6$     &   \onlinecite{yin_electronic_2011s} \\
$U = 3.0$         &                    &   \onlinecite{yin_origin_2008s} \\
$U = 4.0$         &  $J = 0-0.5$       &   \onlinecite{suzuki_microscopic_2013s} \\
\end{tabular}
\end{table}

In Table \ref{tab:literature_u} we present an overview of Hubbard U values used
for calculations for UO$_2$ in the literature. It is remarkable how close most
of these values are to the result of our shell-folded theory.

Nevertheless, a recent work by Koloren\v{c} {\it et
al.}~\cite{kolorenc_electronic_2015s} presented LDA+DMFT calculations using a
larger value of $F^0=6.5$eV.  The authors show that the observed spectral
function closely resembles spectra calculated within the Hubbard I
approximation using a smaller $U$ of 3 eV. Apart from a satellite structure at
about 10 eV, the spectral features between these two theories presented an
impressive match.  They also note that these $U$ values are slightly on the
small side relative to experiment. This is consistent with the observation in
Ref.~\onlinecite{suzuki_microscopic_2013s} that the best description of the
low-temperature ordered phase is obtained within LDA+U with $U = 4$ eV.
Koloren\v{c} {\it et al.}~add an interesting discussion of the differences in
the treatment of screening in LDA+U and LDA+Hubbard I.  They argue that
screening by ligand states is absent at the LDA+U or LDA+Hubbard I levels such
that the $U$ used must already incorporate these screening processes.  For
LDA+DMFT, on the other hand, they argue that some screening comes in through
the coupling of the $f$-states to a the dynamical bath, requiring a larger $U$.
Our cRPA calculations for the reduction of $U$ by screening through the ligands
agrees well with these estimates. 

For completeness we note that there is an additional effect. $U$ values will
depend on the basis used with the two extremes being a set of localized
atomic-like $f$- and $p$-orbitals and a block-diagonal basis where $f$-$p$
hybridization enters effectively. We illustrate this on a simple two-orbital
model with atomic-like $f$- and $p$-orbitals, hybridizing with strength $V$.
The eigenvalues read, in terms of the energies of the local levels $\epsilon_f$
and $\epsilon_p = \epsilon_f - \Delta$  of the $f$ and $p$ state respectively:
\begin{eqnarray}
E = \frac{1}{2}(\epsilon_f + \epsilon_p)
 \pm \frac{1}{2} \sqrt{
(\epsilon_f - \epsilon_p)^2 + 4 |V|^2
}
\end{eqnarray}
and the eigenvectors can be expressed in terms of the atomic
states $| f \rangle$, $| p \rangle$:
\begin{eqnarray}
| \pm \rangle
= \cos(\theta) |f \rangle \pm \sin(\theta) | p \rangle
\end{eqnarray}
with
$\tan(\theta) = -\frac{\Delta}{2 V} + \sqrt{(\frac{\Delta}{2 V})^2 + 1}$.
Since $U^{ff}$ transforms as $\cos(\theta)^4$, the values in an atomic-like
basis are reduced when transforming to a basis that block diagonalizes the $fp$
Hamiltonian. The precise amount of reduction depends on the orbitals since the
hybridization is strongly orbital-dependent.

%%%%%%%%%%%%%%%%%%%%%%%%%%%%%%%%%%%%%%%%%%%%%%%%%%%%%%%%%%%%%%%%%%%%%%%%%%%%%%%%
%%%%%%%%%%%%%%%%%%%%%%%%%%%%%%%%%%%%%%%%%%%%%%%%%%%%%%%%%%%%%%%%%%%%%%%%%%%%%%%%
\section{Relation between cluster calculations and DMFT
calculations}

The rewriting of the Hamiltonian presented in the main text of our paper also
allows for a deeper understanding of the relation between cluster model
calculations and DFT+DMFT. Indeed, by definition, the cluster model assumes the
total particle number to be fixed to specific integer values, and diagonalizes
the cluster Hamiltonian for those. In DMFT, in general, the bath introduces the
possibility of charge fluctuations and the above assumption becomes less
obvious. A cluster approximation for insulators is well justified since the
bath hybridization is small. For metallic systems, restricting the Hamiltonian
the low-energy correlated subspace avoids the problem. The Zhang-Rice-like
construction presented above provides a way to do so explicitly.  Otherwise,
when the O $p$-states are to be kept explicitly in the description, the shifts
induced by the changes in the particle number might have to be taken into
account explicitly.

%%%%%%%%%%%%%%%%%%%%%%%%%%%%%%%%%%%%%%%%%%%%%%%%%%%%%%%%%%%%%%%%%%%%%%%%%%%%%%%%
%%%%%%%%%%%%%%%%%%%%%%%%%%%%%%%%%%%%%%%%%%%%%%%%%%%%%%%%%%%%%%%%%%%%%%%%%%%%%%%%
\bibliographystyle{plain}

\end{document}